\documentclass[runningheads]{llncs}
\usepackage{times}
\usepackage{xcolor}
\usepackage{soul}
\usepackage[utf8]{inputenc}
\usepackage{comment}
\usepackage{color}
\usepackage{framed}
\usepackage[noadjust]{cite}
\usepackage{balance}

\usepackage{epsfig}
\usepackage{graphicx}
\usepackage{amsmath}
\usepackage{amssymb}

\usepackage{url}
\usepackage{multirow}
\usepackage{enumitem}
\usepackage{moresize}
\usepackage{epstopdf}
\usepackage{array}

\usepackage{diagbox}

\usepackage{tcolorbox}

\usepackage{colortbl}
\definecolor{light-gray}{gray}{0.92}
\usepackage{multirow}

\makeatletter
\newcommand{\thickhline}{%
    \noalign {\ifnum 0=`}\fi \hrule height 1pt
    \futurelet \reserved@a \@xhline
}
\newcolumntype{"}{@{\hskip\tabcolsep\vrule width 1pt\hskip\tabcolsep}}
\makeatother

 \let\OLDthebibliography\thebibliography
 \renewcommand\thebibliography[1]{
   \OLDthebibliography{#1}
   \setlength{\parskip}{-1pt}
   \setlength{\itemsep}{0pt plus 0.3ex}
 }

\usepackage{wrapfig}
\usepackage{subfigure}
\usepackage[hang,flushmargin]{footmisc} % remove footnote indentation

\usepackage[font=small,labelfont=bf]{caption} % felix commented, according to template 
\usepackage[belowskip=-15pt,aboveskip=3pt]{caption} % re-arrange tables on last page, select

\usepackage{xspace}
\makeatletter
\DeclareRobustCommand\onedot{\futurelet\@let@token\@onedot}
\def\@onedot{\ifx\@let@token.\else.\null\fi\xspace}
 
\def\ie{\emph{i.e}\onedot}

\def\etal{\emph{et al}\onedot}
\makeatother

\newcommand{\lei}[1]{\textbf{\textcolor{blue}{Lei: #1}}}

\begin{document}

\title{Secure Deep Learning Engineering:\\ 
A Software Quality Assurance Perspective}

\author{Lei Ma\inst{1,2} \and
Felix Juefei-Xu\inst{3} \and
Minhui Xue\inst{4}\and 
Qiang Hu\inst{5} \and
Sen Chen\inst{2} \and\\
Bo Li\inst{6}\and 
Yang Liu\inst{2}\and
Jianjun Zhao\inst{5}\and
Jianxiong Yin\inst{7} \and 
Simon See \inst{7}}

\authorrunning{L. Ma et al.}

\institute{Harbin Institute of Technology, China \and
Nanyang Technological University, Singapore \and Carnegie Mellon University, USA \and  Optus Macquarie University Cyber Security Hub, Australia \and Kyushu University, Japan \and University of Illinois at Urbana–Champaign, USA \and NVIDIA AI Tech Centre, Singapore \\
}

\maketitle

\begin{abstract}
Over the past decades, deep learning (DL) systems have achieved tremendous success and gained great popularity in various applications, such as intelligent machines, image processing, speech processing, and medical diagnostics. 
Deep neural networks are the key driving force behind its recent success, but still seem to be a magic black box lacking interpretability and understanding. This brings up many open safety and security issues with enormous and urgent demands on rigorous methodologies and engineering practice for quality enhancement.
A plethora of studies have shown that the state-of-the-art DL systems suffer from defects and vulnerabilities that can lead to severe loss and tragedies, especially when applied to real-world safety-critical applications.

In this paper, we perform a large-scale study and construct a paper repository of 223 relevant works to the quality assurance, security, and interpretation of deep learning. We, from a software quality assurance perspective, pinpoint challenges and future opportunities towards universal secure deep learning engineering. We hope this work and the accompanied paper repository can pave the path for the software engineering community towards addressing the pressing industrial demand of secure intelligent applications.


\end{abstract}

\section{Introduction}

In company with massive data explosion and powerful computational hardware enhancement, deep learning (DL) has recently achieved substantial strides in cutting-edge intelligent applications, ranging from virtual assistant~(e.g., Alex, Siri), art design~\cite{DBLP:journals/corr/ElgammalLEM17}, autonomous vehicles~\cite{7410669, Eliot:2017:AAA:3165260}, to medical diagnoses~\cite{dl_eye, dl_cancer} -- tasks that until a few years ago could be done only by humans.
DL has become the innovation driving force of many next generation's technologies. We have been witnessing on the increasing trend of industry stakeholders' continuous investment on DL based intelligent system~\cite{google_invest, jaguar_invest, gm_invest, honda_invest, toyota_softbank}, penetrating almost every application domain, revolutionizing industry manufacturing as well as reshaping our daily life.

However, current DL system development still lacks systematic engineering guidance, quality assurance standards, as well as mature toolchain support.
The \emph{magic box}, such as DL training procedure and logic encoding~(as high dimensional weight matrices and complex neural network structures), further poses challenges to interpret and understand behaviors of derived DL systems~\cite{2017arXiv170208608D, 2017arXiv171111279K, trust_ai}. The latent software quality and security issues of current DL systems, already started emerging out as the major vendors, rush in pushing products with higher intelligence~(e.g., Google/Uber car accident~\cite{google_crash, uber_crash}, Alexa and Siri could be manipulated with hidden command~\cite{alex_siri}. A DL image classifier with high test accuracy is easily fooled by a single pixel perturbation~\cite{one_pixel}). Deploying such cocooned DL systems to real-world environment without quality and security assurance leaves high risks, where newly evolved cyber- and adversarial-attacks are inevitable.

To bridge the pressing industry demand and future research directions, this paper performs a large-scale empirical study on the most-recent curated 223 relevant works on deep learning engineering from a software quality assurance perspective. 
Based on this, we perform a quantitative and qualitative analysis to identify the common issues that the current research community most dedicated to.
With an in-depth investigation on current works, and our in-company DL development experience obtained, we find that the development of secure and high quality deep learning systems requires enormous engineering effort, while most AI communities focus on the theoretical or algorithmic perspective of deep learning.
Indeed, the development of modern complex deep learning systematic solutions could be a challenge for an individual research community alone.
We propose the \emph{Secure Deep Learning Engineering} (SDLE) development process specialized for DL software, 
which we believe is an interdisciplinary future direction~(e.g., AI, SE, security) towards constructing DL applications, in a systematic method from theoretical foundations, software \& system engineering, to security guarantees.
% This work takes the first step towards such a direction.
We further discuss current challenges and opportunities in SDLE from a software quality assurance perspective. 

To the best of our knowledge, our work is the first study to vision SDLE, from the quality assurance perspective, accompanied with a state-of-the-art literature curation.
We hope this work facilitates drawing attention of the software engineering community on necessity and demands of quality assurance for SDLE, which altogether lays down the foundations and conquers technical barriers towards constructing robust and high quality DL applications. 
The repository website is available at:
\begin{center}
\footnotesize{\url{https://sdle2018.github.io/SDLE/V1.1/en/Home.html}
}
    
\end{center}

\section{Research Methodology}
This section shows the research questions, and discusses the detail of paper collection procedure.

\subsection{Research Questions}
This paper mainly focuses on following research questions.
\begin{itemize}[wide=1pt,leftmargin=10pt]

    \item {\bf RQ1:} What are mostly research topics and the common challenges relevant to quality assurance of deep learning? 
    
    \item {\bf RQ2:} What is secure deep learning engineering and its future direction in perspective of quality assurance?

\end{itemize}

The RQ 1 identifies the mostly concerned topics in the research community and their common challenges, while RQ2 concerns the key activities in SDLE life cycle, based on which we discuss our vision and future opportunities.

\subsection{Data Collection Methodology}
Extensive research contributions are made on deep learning over the past decades, we adopt the following procedure to select works most relevant to the theme of our paper repository.

\begin{itemize}[wide=1pt,leftmargin=10pt]
    \item We first collect papers from conferences listed on the \emph{Computer Science Rankings} within the scope of AI \& machine learning, software engineering, and security.\footnote{\scriptsize\url{http://csrankings.org/#/index?all}}
    To automate the paper collection procedure, we develop a Python-based crawler to extract paper information of each listed conference since the year 2000 and filter with keywords.
    \item To further reduce the search space for relevant topics, we use keywords~(e.g., deep learning, AI, security, testing, verification, quality, robustness) to filter the collected papers.
    \item Even though, scraping all the listed conferences may still be insufficient, we therefore crawl outwards -- extract all the related work for each keyword-filtered paper and crawl one level down of these papers.
    \item This finally results in 223 papers and we manually confirmed and labeled each paper to form a final categorized list of literature.
\end{itemize}

\begin{figure}[t]
  \centering
  \includegraphics[width=1.0\textwidth]{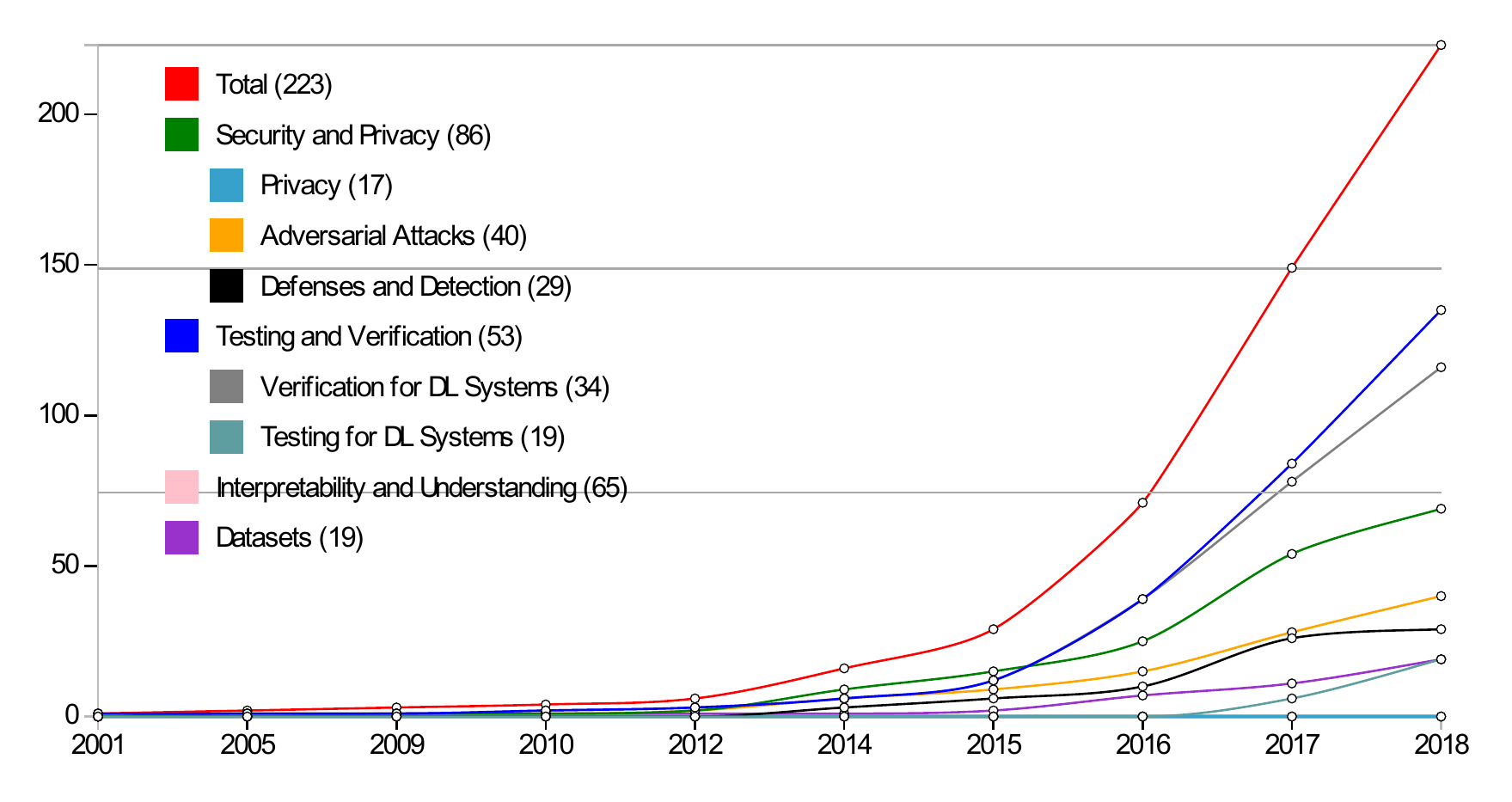}
  \caption{The accumulative number of selected publications over years}
  \label{fig:chart}
  \vspace{0.1cm}
\end{figure}

\noindent \textbf{Paper Category and Labeling.} 
To categorize the selected papers, we perform paper clustering by taking into account the title, abstract, and listed keywords. Based on further discussion of all authors~(from both academia and industry with AI, SE, and security background), we eventually identify four main paper categories, and seven fine-grained categories in total~(see Figure~\ref{fig:chart}).
In the next step, three of the authors manually label each paper into a target category independently, and discuss the non-consensus cases until an agreement is reached.

 \begin{figure}[t]
   \centering
   \includegraphics[width=0.95\textwidth]{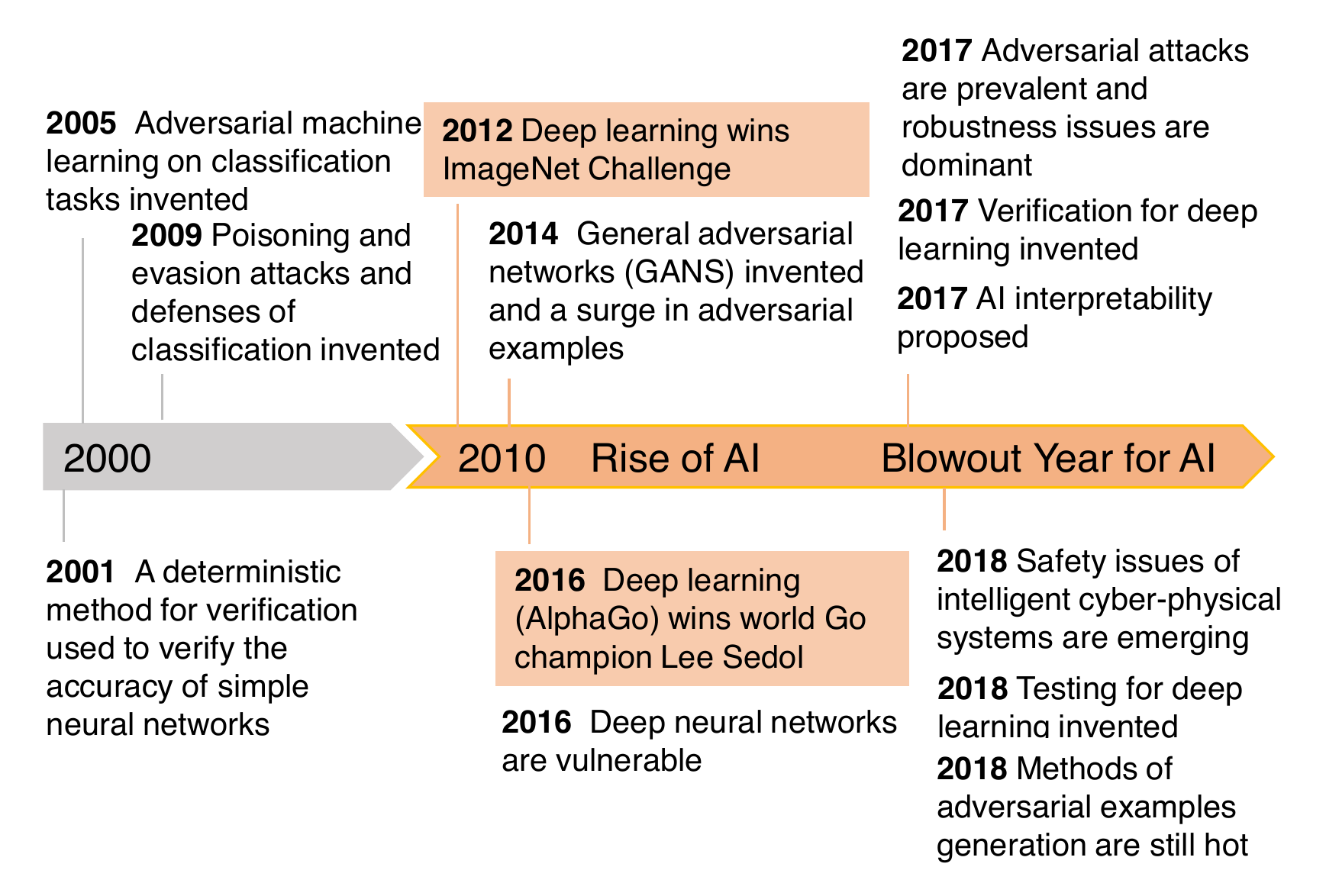}
   \caption{Milestones of deep learning engineering relevant to security and software quality. }
   \label{fig:timeline}
     \vspace{0.1cm}
 \end{figure}

\noindent \textbf{The Dataset and the Trend.} 
Figure~\ref{fig:chart} shows the general trends of publication on secure deep learning research area, where the publication number~(i.e., both total paper as well as in each category) dramatically increases over years.
Such booming trend becomes even more obvious accompanied with the milestones of DLs~(e.g., DL won ImageNet Challenge in 2012, AlphaGo defeated human championship in 2016), which is highlighted in Fig.~\ref{fig:timeline}.
For the four main categories, we find the most publications are relevant to Security and Privacy~(SP, 86 papers), followed by Interpretability and Understanding~(IU, 65), Testing and Verification~(TV, 53), and Datasets~(17). 

The SP category with the highest paper publication number is not surprising. Since Goodfellow et al.~\cite{goodfellow2014explaining} posted the security issues of DLs, it attracted both the AI and security communities to escalate and burst a research competition on defending and attacking techniques.
Even though, it still lacks a complete understanding on why current DL systems are still vulnerable against adversarial attacks. This draws the attention of researchers on interpreting and understanding how DL works, which would be important for both application and construction of robust DLs. As the recent emerging investment blowout in DL applications to safety-critical scenarios~(e.g., autonomous driving, medical diagnose), its software quality has become a big concern, where researchers find that the different programming paradigm of DL makes existing testing and verification techniques unable to directly handle DLs~\cite{pei2017deepxplore,huang2017safety, ma2018deepgauge,DBLP:journals/corr/abs-1805-00089}. Therefore, we have observed that many recent works are proposing novel testing and verification techniques for DLs, from testing criteria, test generation techniques, test data quality evaluation, to static analysis. Meanwhile, the dataset benchmarks of different DL application domains emerge to grow as well~\cite{xiang2016objectnet3d,2018arXiv180306184H, Ramanishka_2018_CVPR, Chen_2018_CVPR}, in order to facilitate the study of solving domain-specific problems by DLs~(e.g., image classification, 3D object recognition, autonomous driving, skin disease classification).

\noindent \textbf{Common Issues.} 
In contrast to traditional software of which the decision logic is mostly programmed by human developers, deep learning adopts a data-driven programming paradigms. Specifically, a DL developer's major effort is to prepare the training data~(including knowledge to resolve a task) and neural network architecture, after which the decision logic is automatically obtained through the training procedure. On one hand, this paradigm reduces the burden of a developer who manually crafts the decision logic. On the other hand, for a DL developer, the logic training procedure is almost like a magic-box driven by an optimization technique.
Due to the decision logic of DL is encoded into a DNN with high dimensional matrices, the interpretation and understanding, training procedure, as well as the obtained decision logic are all very difficult~\cite{DBLP:journals/corr/Lipton16a}, which could be a root cause and a common challenge among all categories.
For example, without completely understanding the decision logic of DL, it is hard to know in what particular case an adversarial attack could penetrate, and how we could defend against such attacks. In the case of testing, extensive studies are performed on analysis of traditional software bugs, their relations to software development activities, as well as techniques for defect detection. However, a comprehensive empirical study and understanding on why DL bugs occur still could not be well explained, let alone the root case analysis.

\begin{figure}[t]
  \centering
  \includegraphics[width=1.0\textwidth]{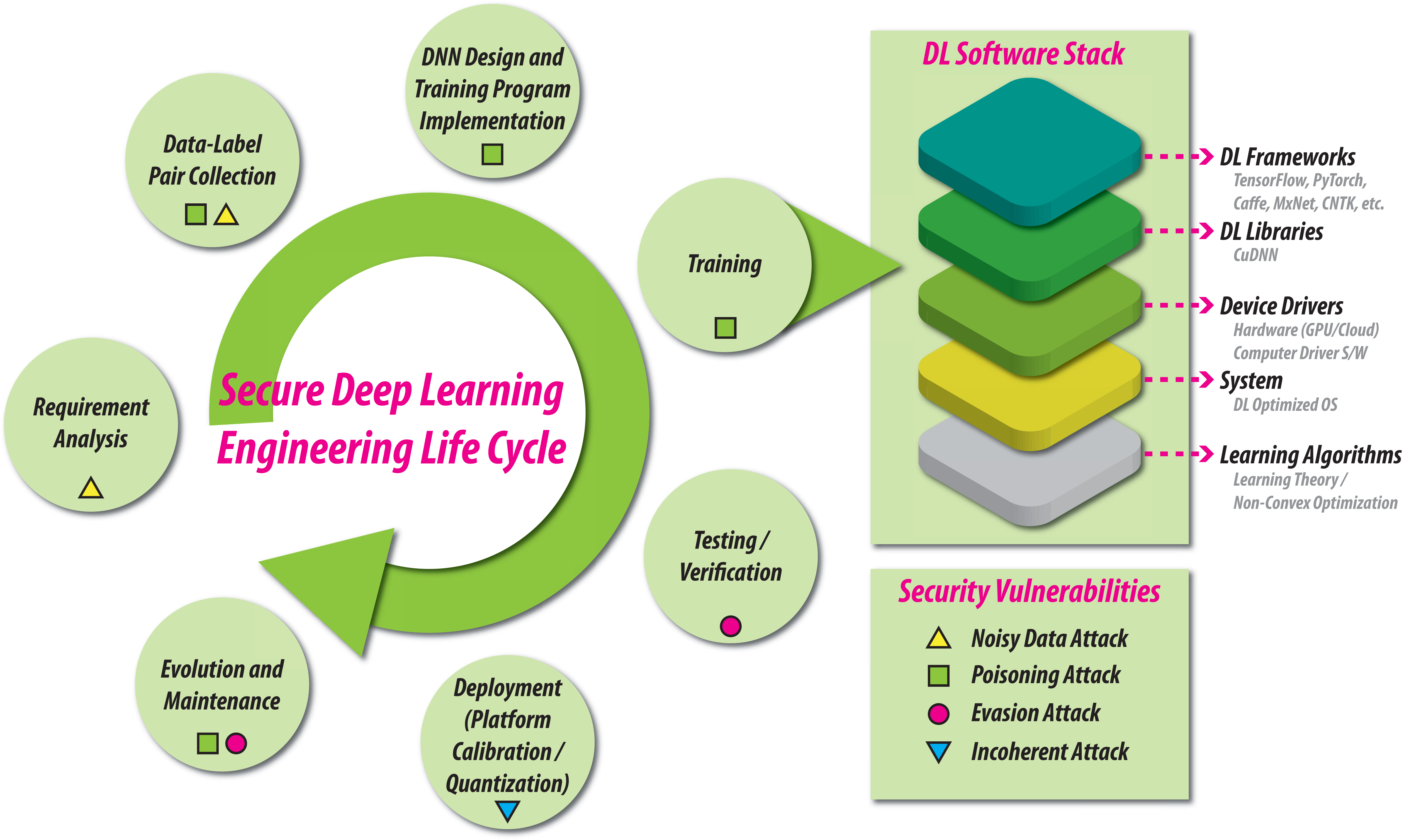}
  \caption{Secure deep learning engineering life cycle}
  \label{fig:sdle_lc}
    % \vspace{0.1cm}
\end{figure}

\section{Secure Deep Learning Engineering Life Cycle}

Due to the fundamental different programming paradigms of deep learning and traditional software, the secure deep learning engineering practice and techniques are largely different with traditional software engineering, although the major development phases could still be shared. 

\begin{tcolorbox}[colframe=black, colback=white]
	We define \emph{Secure Deep Learning Engineering (SDLE) as an engineering discipline of deep learning software production, through a systematic application of knowledge, methodology, practice on deep learning, software engineering and security, to requirement analysis, design, implementation, testing, deployment, and maintenance of deep learning software.}
\end{tcolorbox}

Figure~\ref{fig:sdle_lc} shows the key development phases of SDLE. In the rest of this section, we first describe each of the key development phases, their uniqueness and difference compared with traditional practices in software engineering, and then we discuss the security issues in current SDLE.
In the next section, we explain the quality assurance necessity in SDLE life cycle, and highlight the challenges and opportunities.

\noindent \textbf{Requirement Analysis.} Requirement analysis investigates the needs, determines, and creates detailed functional documents for the DL products.
DL-based software decision logic is learned from the training data and generalized to the testing data. Therefore, the requirement is usually measured in terms of an expected prediction performance, which is often a statistics-based requirement, as opposed to the rule-based one in traditional SE.

\noindent \textbf{Data-Label Pair Collection.} After the requirements of the DL software become available, a DL developer~(potentially with domain experts for supervision and labeling) tries to collect representative data that incorporate the knowledge on the specific target task. For traditional software, a human developer needs to understand the specific task, figures out a set of algorithmic operations to solve the task, and programs such operations in the form of source code for execution. On the other hand, one of the most important sources of DL software is training data, where the DL software automatically distills the computational solutions of a specific task. 

\noindent \textbf{DNN Design and Training Program Implementation.} When the training data become available, a DL developer designs the DNN architecture, taking into account of requirement, data complexity, as well as the problem domain. For example, when addressing a general purpose image processing task, convolutional layer components are often included in the DNN model design, while recurrent layers are often used to process natural language tasks.
To concretely implement the desired DNN architecture, a DL developer often leverages an existing DL framework to encode the designed DNN into a training program. Furthermore, he needs to specify the runtime training behaviors through the APIs provided by the DL framework~(e.g., training epochs, learning rate, GPU/CPU configurations). 

\noindent \textbf{Runtime Training.} After the DL programming ingredients~(i.e., training data and training program) are ready. The runtime training procedure starts and systematically evolves the decision logic learning towards effectively resolving a target task.
The training procedure and training program adjustment might go back-and-forth several rounds until a satisfying performance is achieved.
Although the training program itself is often written as traditional software~(e.g., in \texttt{Python}, \texttt{Java}), the obtained DL software is often encoded in a DNN model, consisting of the DNN architecture and weight matrices. 
The training process plays a central role in the DL software learning, to distill knowledge and solution from the sources. It involves quite a lot of software and system engineering effort to realize the learning theory to DL software~(see Figure~\ref{fig:sdle_lc}) over years.

\noindent \textbf{Testing \& Verification.}
When the DNN model completes training with its decision logic determined, it goes through the systematic evaluation of its generality and quality through testing~(or verification). Note that the testing activity in the AI community mainly considers whether the obtained DL model generalizes to the prepared test dataset, to obtain high test accuracy. 
On the other hand, the testing activity~(or verification) in SDLE considers a more general evaluation scope, such as generality, robustness, defect detection, as well as other nonfunctional requirement~(e.g., efficiency). The early weakness detection of the DL software provides valuable feedback to a DL developer for solution enhancement.

\noindent \textbf{Deployment.}
A DL software passed the testing phase reaches a certain level of quality standard, and is ready to be deployed to a target platform. 
However, due to the platform diversity, DL framework supportability, and computation limitations of a target device, the DL software often needs to go through the platform calibration~(e.g., compression, quantization, DL framework migration) procedure for deployment on a target platform. For example, once a DL software is trained and obtained on the \texttt{Tensorflow} framework, it needs to be successfully transformed to its counterpart of \texttt{TensorflowLite}~(resp. \texttt{CoreML}) framework to \texttt{Android}~(resp. \texttt{iOS}) platform. It still needs to go through on device testing after deployment, and we omit the testing phase after deployment for simplicity.

\noindent \textbf{Evolution and Maintenance.}
After a DL product is deployed, it might experience the procedure of modification for bug correction, performance and feature enhancements, or other attributes. 
The major effort in evolution and maintenance phases relies on the manually revision on design, source code, documentation, or other software artifacts. On the other hand, DL software focuses more on  comprehensive data collection, DL model continuous learning~(e.g., re-fitting, retro-fitting, fine tuning, and re-engineering).

\noindent\textbf{Security Issues in DL.}
The current practice of security in deep learning has fallen into the trap that many other domains have experienced. Almost every month new attacks are identified~\cite{biggio2013evasion,goodfellow2014explaining,papernot2016limitations,moosavi2015deepfool,carlini2017towards,xu2016automatically,chen2017ead}
followed by new countermeasures~\cite{papernot2015distillation,xu2017feature} which are subsequently broken~\cite{carlini2017towards,he2017adversarial}, and so on ad-infinitum.  There is a broad and pressing need for a frontier-level effort on trustworthiness and security in DL to break this cycle of attacks and defenses.  We have a unique opportunity at this time---before deep learning is widely deployed in critical systems---to
develop the theory and practice needed for
robust learning algorithms that provide rigorous and meaningful guarantees.
If we rethink the SDLE life cycle~(see Figure~\ref{fig:sdle_lc}), security vulnerabilities can happen in almost every step. For instance, for the training related steps such as \emph{Requirement Analysis, Data-Label Pair Collection} and \emph{DNN design and training}, poisoning attacks can easily happen via manipulating training data. In the testing related steps, such as \emph{testing \& verification}and \emph{deployment}, evasion attacks can take place by perturbing the testing data slightly (e.g. adversarial examples). In addition, when deploying the DL software to different platforms or with different implementation frameworks, there will always be opportunities for adversaries to generate attacks from one to the other.

We believe many of  these security issues are highly intertwined the quality of current DL software, lacking systematic quality assurance solutions over the entire SDLE process which is largely missed in research works as described in the next section.

\section{Towards Future Quality Assurance of SDLE}

Over the past decades, software quality assurance discipline \cite{Pressman:2009:SEP:1593949,Ruparelia:2010:SDL:1764810.1764814} has been well-established for traditional software, with many experiences and practices widely applied in software industry.  
However, the fundamentally different programming paradigm and decision logic representation of DL software make existing quality assurance techniques unable to be directly applied, forcing us to renovate the entire quality assurance procedure for SDLE. In this section, we pose our vision and challenges on quality assurance in SDLE to guide future research. 

From the very beginning of SDLE, we need to rethink how to accurately define, specify, and document the of DL software requirement, especially for the functional requirements. This leaves us a question whether we should follow a statistical based approach, a rule based approach, or their combination, which has not been well investigated yet.

The training data play a key role in shaping the learning process and DL decision logic. However, most current research treats the training data as high quality for granted, without a systematic quality control, inspection and evaluation process. As poisoning attacks show, many incorrect behaviors and security issues could be introduced with the maliciously tweaked training data. How to select the suitable size while representative data would be an important question. In addition, data supervision and labeling process is also labor intensive and error prone. For example, ImageNet dataset contains more than one million general purpose images. We also need to provide assistance and quality control for the labeling procedure.

It becomes even more challenging, when it comes to the training program implementation and runtime training. Most state-of-the-art DL frameworks are implemented as traditional software on top of the DL software stack. Even the learning theory is perfect, it still has a big gap to transfer such ideally designed DL models to a DL application encoded on top of the DL framework. One big challenge is how to ensure the software stack~(e.g., hardware drivers, DL library, DL framework) correctly implements the learning algorithm. Another challenge is to provide useful interactive support of the training process. Most current DL training procedure only shows training loss~(accuracy), validation loss~(accuracy), which is mostly a black box to a DL developer. When, the training procedure goes beyond expectation, the root cause analysis becomes extremely difficult, which may come from the DL architecture issue, training program implementation issue, or the hardware configuration issue. Hence, the software engineering community needs to provide the novel debugging, runtime monitoring, and profiling support for the training procedure, which is involved with non-determinism and runtime properties hard to specify.

The large input space has already been a challenge for testing and verifying traditional software. 
Such a challenge is further escalated for DL software, due to its high dimensional input space and the internal latent space. Even though, traditional software testing has already explored many testing criteria as the goal to guide testing. How to design a suitable testing criteria to capture the testing confidence still remains unclear. 
Even with some preliminary progress on testing criteria designed for DLs~\cite{pei2017deepxplore, ma2018deepgauge, DBLP:journals/corr/abs-1805-00089, criteria_guiding}, there are many more testing issues needed to be addressed, such as how to effectively generate tests~\cite{2018arXiv180901266X, ma2018combinatorial}, how to measure the test data quality~\cite{ma2018deepmutation}, and how to test DL robustness and vulnerabilities~\cite{ 2018arXiv180902444C,2018arXiv180605859B}.

Further DL challenge comes up with current deployment process: (1) target device computation limitations, and (2) DL framework compatibility across platforms. The DL software is mostly developed and trained on severs or PCs with GPU support. When it needs to be deployed on a mobile device or edge device with limited computation power, the DL software must be compressed or quantized for computation/energy efficiency, which could introduce defects. How to ensure the quality and detect the potential issues during this process is an important problem.
In addition, the current DL frameworks might not always be  supported by different platforms. For example, the Tensorflow is not directly supported by Android or iOS, and how to make DL software cross-platform compatible would be an important direction. Finally the quality assurance concerns in DL software evolution and maintenance are mostly focused on avoiding introducing defects during change, which might rely on regression testing. However, how to effectively evolve the DL software still remains unknown, which we leave as an open question for further study.

\section{Conclusion}

Considering deep learning is likely to be one of the most transformative technologies in the 21st century, it appears essential that we begin to think how to design fully fledged deep learning systems under a well-tested development discipline. This paper defines the secure deep learning engineering and discusses the current challenges, opportunities, and puts forward open questions from the quality assurance perspective, accompanied with a paper repository. We hope our work can inspire future studies towards constructing robust and high quality DL software.

\bibliographystyle{splncs04}
\bibliography{sample-bibliography}

\end{document}